\documentclass{elsart}
\input axodraw1.sty

\def\s{\sigma} 
\def\G{\Gamma} 
\def\an{analytic}
\def\ca{\an{} continuation}

\def\hf{hypergeometric functions}
\def\ndim{NDIM}
\def\yy{photon-photon scattering}

\begin{document}
\begin{frontmatter}
\title{{\bf NEGATIVE DIMENSIONAL INTEGRATION FOR MASSIVE FOUR POINT
FUNCTIONS--I: THE STANDARD SOLUTIONS }} 
\author{Alfredo T. Suzuki and Alexandre G. M. Schmidt}
\address{Instituto de F\'{\i}sica Te\'orica, Universidade Estadual
Paulista, R.Pamplona, 145, S\~ao Paulo  SP, CEP 01405-900, Brazil }

\begin{abstract}
Feynman diagrams are the best tool we have to study perturbative 
quantum field theory. For this very reason the development of any new 
technique which allows us to compute Feynman integrals is welcome. 
By the middle of the 80's, Halliday and Ricotta suggested the 
possibility of using negative dimensional integrals to tackle the 
problem. The aim of this work is to revisit the technique as such and 
check up on its possibilities. For this purpose, we take a box diagram 
integral contributing to the photon-photon scattering amplitude in 
quantum electrodynamics using the negative dimensional integration 
method. The reason for this choice of ours is twofold: Firstly, it is 
a well-studied integral with well-known results, and secondly because 
it bears in its integrand the complexities associated with four massive
propagators of the intermediate states.
\end{abstract}

\end{frontmatter}

\section{Introduction.}

Scattering amplitudes, radiative corrections, $\beta $ functions of
renormalization group, etc., all require the computation of Feynman 
integrals\cite{peskin,zuber}, which are the more complex to evaluate
the more loops in a given diagram one has. Since this approach is still
the best technique we have to study quantum field theory (QFT)
perturbatively, solving Feynman integrals becomes basic to do any
serious study on physical processes of interest involving those
quantities. Such computations become harder to do not only with
increasing number of loops, but also with increasing number of massive
particles in the intermediate states. This subject becomes even more 
compeling as we realize that increasingly higher energies are
disposable at particle accelerators and tests for the standard
electroweak theory are pushed to new limits. It is then clear that in
order to study radiative corrections there will unavoidably lead us to
the necessity of dealing with Feynman integrals containing massive
intermediate vector bosons, and our ability to 
perform them will play a key role.

The standard way of solving such integrals starts with the introduction 
of Feynman parameters, Wick rotation and then finally, integration. 
This method is somewhat tedious and sometimes it is not possible to
solve exactly the parametric integrals. For this reason physicists
developed several other techniques to calculate Feynman
integrals\cite{boos}. A technique known as negative dimensional
integration method (NDIM) \cite{halliday,suzuki2,suzuki1} has also been
considered to tackle the problem. 

Our aim in this work is to further check up on NDIM as a useful tool
for the referred task. For this purpose, we use NDIM to evaluate the
box diagram contributing to the \yy{} (see fig.1) in quantum
electrodynamics (QED). The outline for our paper is as follows: In
section 2 we give a brief review of the methodology to be employed,
while in section 3 we compute the integral proper writing down the two
well-known hypergeometric series representations for it. In section 4
we consider two particular cases of the given integral, namely,
integrals with three and two massive propagators respectively. Finally,
in the last section, we make a few comments about the six new results
we have for this diagram, which will be the subject to be addressed in
our shortly forthcoming paper. Also we mention more complicated
integrals, like the ones arising in one-loop correction to Bhabha
scattering in QED, and off-shell two-loop three point
graphs\cite{twoloops}. 

{\small {\ }}
\begin{figure}
\begin{center}
\vspace{25mm}
{\small 
\begin{picture}(400,250)(0,-10)
\thicklines
\Photon(90,250)(150,200)5 5
\Photon(90,90)(150,140)5 5
\ArrowLine(150,140)(150,200)
\ArrowLine(250,140)(150,140)
\ArrowLine(250,200)(250,140)
\ArrowLine(150,200)(250,200)
\Photon(250,200)(310,250)5 5
\Photon(250,140)(310,90)5 5

\thinlines
\put(110,250){\vector(1,-1){30}}
\put(110,90){\vector(1,1){30}}
\put(270,200){\vector(1,1){30}}
\put(270,140){\vector(1,-1){30}}

\Text(80,70)[c]{$p-k_1$} 
\Text(90,270)[c]{$k_1$}
\Text(320,70)[c]{$p-k_2$}
\Text(320,270)[c]{$k_2$}

\Text(120,170)[c]{$q-k_1$} 
\Text(200,220)[c]{$q$}
\Text(280,170)[c]{$q-k_2$}
\Text(200,120)[c]{$q-p$}
\end{picture}
}
\end{center}
\caption{Feynman diagram for \yy{} in the s-channel}
\end{figure}

\section{Integration in Negative Dimensions.}

NDIM was introduced by Halliday and Ricotta\cite{halliday} some years 
ago. In this section we present a brief review of this technique. 
Basically what one does is to perform an analytic continuation

\begin{equation}
\int \frac{\d^Dq}{(A)(B)(C)...}\;\;\longrightarrow
^{\!\!\!\!\!\!\!\!\!\!\!\!AC}\;\;\int \d^Dq(A)(B)(C)...
\end{equation}
so that one gets a polynomial integral in $D<0$\footnote{The negative
dimensional linear operator object 
here can be seen from the viewpoint of positive dimensional fermionic
integration\cite{halliday2}.} from a rather 
complicated one in $D>0$. We then solve it in $D<0$ and go back
\cite{suzuki2} to $D>0$, through another analytic continuation. One of 
the advantages of NDIM is that simultaneously we get several 
hypergeometric series representations for the integral in $D>0$, i.e., 
we obtain expressions for all the possible regions, physical and 
non-physical alike, of the external momenta.

We start from the relation\cite{halliday,suzuki2,suzuki1} between a 
gaussian integral and its counterpart in negative dimensions

\begin{equation}
\int \d^D\!q\exp {(-\lambda q^2)}=\left( \frac \pi \lambda \right)
^{D/2}=\sum_{j=0}^\infty \frac{(-1)^j\lambda ^j}{j!}\int \d^Dq(q^2)^j
\end{equation}
where in the last step we have expanded the exponential function in 
Taylor series. Just like in dimensional regularization\cite{thooft} we 
take this expression as the definition of the negative $D-$dimensional
integral 
\cite{nash}. The middle term is an analytic function of $D$ so the 
integral on the right hand side is also an analytic function of $D$
\cite{marku,ww,eden}.

From this equation we get,

\begin{equation}
\int \d^D\!q(q^2)^j=(-1)^j\pi ^{D/2}\delta _{D/2,-j}\Gamma (1+j)
\end{equation}

In a similar way, we can solve, e.g.,

\begin{eqnarray}  \label{J}
J(i,j,k,l;m) &=& \int \d^D\! q\; \left(q^2-m^2\right)^i
\left[(q-p)^2-m^2\right]^j\left[(q-k_1)^2-m^2\right]^k\nonumber\\
&& \times\left[(q-k_2)^2-m^2\right]^l
\end{eqnarray}
whose counterpart in $D>0$ is the integral

\begin{eqnarray}
K(i,j,k,l;m)&=&\int \frac{\d^D\!q\;}{\left( q^2-m^2\right) ^i\left[
(q-p)^2-m^2\right] ^j\left[ (q-k_1)^2-m^2\right] ^k} \nonumber\\
&& \times\frac{1}{\left[(q-k_2)^2-m^2\right] ^l}  \label{K}
\end{eqnarray} 

This is one of the integrals that contributes to the \yy{} amplitude in 
QED and it is the one we want to evaluate in our ``lab test'' for NDIM. 
Of course, since the external photons are real particles, they are 
on-shell, i.e., we consider here that
$k_1^2=k_2^2=(p-k_1)^2=(p-k_2)^2=0$ (see fig.1).

So, to begin with, let our ``gaussian integral'' be
\begin{eqnarray}
I &=&\int \d^D\!q\exp {\left( -\alpha (q^2-m^2)-\beta \left[
(q-p)^2-m^2\right] -\gamma \left[ (q-k_1)^2-m^2\right] -\right. }  
\nonumber\label{gauss} \\
&&-\omega \left. \left[ (q-k_2)^2-m^2\right] \right)\end{eqnarray}

Completing the square, integrating over $q$ and expanding the 
exponential, we get

\begin{eqnarray}
I &=&\pi ^{D/2}\sum_{n_i=0}^\infty \frac{(-s)^{n_1}(-t)^{n_2}(-m^2)^
{n_3}\alpha ^{n_2+n_4}\beta^{n_2+n_5}\gamma^{n_1+n_6}\omega^{n_1+n_7}}
{n_1!n_2!n_3!n_4!n_5!n_6!n_7!}  \nonumber  \label{geral} \\
&&\times \Gamma (1+n_3-n_1-n_2-\frac{D}{2})
\end{eqnarray}
where $s$ and $t$ are the Mandelstam variables (see fig.1) and $m$ here
is the mass of the virtual matter fields. Since we use a multinomial
expansion the sum index above must satisfy the constraint

\[ n_4+n_5+n_6+n_7 = n_3-n_1-n_2-\frac{D}{2} \]

On the other hand, expanding the exponential of (\ref{gauss}), we have

\begin{equation}
I=\sum_{i,j,k,l=0}^\infty \frac{(-1)^{i+j+k+l}\alpha ^i\beta ^j\gamma
^k\omega ^l}{\Gamma (1+i)\Gamma (1+j)\Gamma (1+k)\Gamma (1+l)}
J(i,j,k,l;m)\label{gauss2}\end{equation}
and comparing the expressions (\ref{geral}) and (\ref{gauss2}) we
obtain a general relation for the integral $J(i,j,k,l;m)$ and a system
of linear algebraic  
equations linking the sum indices $n_i$, with five equations 
and seven variables, i.e., we have twenty-one distinct solutions for 
this system. So, \ndim{} transfers the problem of calculating Feynman 
integrals to solving systems of linear algebraic equations, which is an
easier task to perform, of course. 

And for the particular system we are dealing with above, there are six
trivial solutions, which are of no interest at all and therefore
discarded, while five are of the hypergeometric type, which will give
the known results. The other ten solutions --- the new results --- will
be the subject of our shortly forthcoming paper.

\section{Hypergeometric Series Representations.}

Solving the system we find five hypergeometric series which can be
divided into two sets according to its variables, $\{I_1\}$, and
$\{I_2, I_3, I_4, I_5\}$. The solution in the first category is

\begin{eqnarray}
{  I}_1 &=&\left( \frac{-\pi }2\right) ^{D/2}\frac{2\sqrt{\pi }%
(-2m^2)^\sigma\Gamma (-\sigma )}{\Gamma (\frac{1}{2}-\frac{\s}{2}+
\frac{D}{4}) \Gamma (-\frac{\s}{2}+\frac{D}{4})}\sum_{n_1,n_2=0}^\infty
\left( \frac s{4m^2}\right) ^{n_1}  \label{I1}\\
&&\times \left( \frac t{4m^2}\right) ^{n_2}\frac{%
(-i|n_1)(-j|n_1)(-k|n_2)(-l|n_2)(-\sigma |n_1+n_2)}{n_1!n_2!(-\frac{\s}
{2}+\frac{D}{4}|n_1+n_2)
(\frac{1}{2}-\frac{\s}{2}+\frac{D}{4}|n_1+n_2)}  
\nonumber\end{eqnarray}

where we define $\sigma=i+j+k+l+\frac{D}{2}$ and use the Pochhammer
symbol 
\[
 (a|k)\equiv (a)_k=\frac{\Gamma (a+k)}{\Gamma (a)} .
\]

Substituting $i=j=k=l=-1$, we get the integral (\ref{K}) with exponents
corresponding to the one we want to calculate for box diagrams. Then,
the first hypergeometric series representation yields (in four
dimensions) 

\begin{equation}  \label{J1}
J_1(-1,-1,-1,-1;m)=\frac{\pi^2}{6m^4}F_3\left(1,1,1,1;\frac{5}{2}\left|
\frac{s}{4m^2},\frac{t}{4m^2}\right)\right.
\end{equation}

This result is exactly what was already obtained by
Davydychev\cite{stand} using the Mellin-Barnes' representation for
massive propagators\cite{boos}. Note that since we are in Euclidean
space there is an overall factor $i$ difference when 
compared to Davydychev's result, obtained in Minkowski space (his
result has the extra factor $i$). This expression is symmetric in $s$
and $t$ and is nonvanishing for $s=t=0$. It will allow us, in section
4, to read off the particular cases where the integral has three and
two propagators respectively, and also the \ca{} to other regions of
external momenta. This expression is valid in the region of convergence
of the series which defines the $F_3$ hypergeometric
function\cite{bateman,appel}, 

\begin{equation}
F_3(\alpha ,\alpha ^{\prime },\beta ,\beta ^{\prime };\gamma
|x,y)=\sum_{j,k=0}^\infty \frac{x^jy^k}{j!k!}\frac{(\alpha |j)(\alpha
^{\prime }|k)(\beta |j)(\beta ^{\prime }|k)}{(\gamma |j+k)}
\end{equation}
where $|x|<1$ and $|y|<1$. In other words, it is valid below the
threshold of pair production. It is also suitable for studying the
non-relativistic limit since the Mandelstam variable $s$ must be less
than $4m^2$. We note that $s=4m^2$ defines the point where the process
changes its nature, that is, there 
exists the possibility of pair creation and this fact manifests itself
in the amplitude as a branch point in the Feynman
integral\cite{zuber,eden}.  

The hypergeometric function in (\ref{J1}) can be expressed in terms of
its double integral representation\cite{bateman}, and from there one
can arrive at the standard result expressed in terms of a rather
cumbersome sum of logarithms and dilogarithms\cite{stand,tollis}.

The next set of solutions, also obtained by Davydychev\cite{stand}, has
variables $4m^2/s$ and $4m^2/t$, the inverse of the ones in the first
solution. In the following we write down these four solutions:
 
\begin{eqnarray}
{ I}_2 &=& \frac{2\pi^{D/2}(-t)^l(-s)^j(-m^2)^{D/2+i+k}(-i|j)(-k|l)}
{(1+i-j+k-l|\frac{D}{2}+j+l)}\sum_{n_1,n_2=0}^\infty \frac{1}{n_1!n_2!} 
\left(\frac{4m^2}{s}\right)^{n_1}  \nonumber \\
&&\times\left(\frac{4m^2}{t}\right)^{n_2} \frac{%
(-j|n_1)(-l|n_2)\left(\left.\frac{1+i-j+k-l}{2}\right|n_1+n_2\right)}
{(1+i-j|n_1)(1+k-l|n_2)(1+i+k+\frac{D}{2}|n_1+n_2)}\nonumber \\
&&\times\left(\left.1+\half (i-j+k-l)\right|n_1+n_2\right),  
\end{eqnarray}

\begin{eqnarray}
{  I}_3 &=& \frac{2\pi^{D/2}(-t)^k(-s)^j(-m^2)^{\frac{D}{2}+i+l}(-i|j)
(-l|k)}{(1+i+j-k-l|\frac{D}{2}+j+k)}\sum_{n_1,n_2=0}^\infty\frac{1}
{n_1!n_2!} \left(\frac{4m^2}{s}\right)^{n_1}  \nonumber \\
&&\times \left(\frac{4m^2}{t}\right)^{n_2}\frac{%
(-j|n_1)(-k|n_2)\left(\left.\frac{1+i-j-k+l}{2}\right|n_1+n_2\right)}
{(1+i-j|n_1)(1-k+l|n_2)(1+i+l+\frac{D}{2}|n_1+n_2)}\nonumber \\
&&\times\left(\left.1+\half (i-j-k+l)\right|n_1+n_2\right),  
\end{eqnarray}

\begin{eqnarray}
{  I}_4 &=& \frac{2\pi^{D/2}(-t)^k(-s)^i(-m^2)^{\frac{D}{2}+j+l}(-j|i)
(-l|k)}{(1-i+j-k+l|\frac{D}{2}+i+k)}\sum_{n_1,n_2=0}^\infty \frac{1}
{n_1!n_2!} \left(\frac{4m^2}{s}\right)^{n_1}  \nonumber \\
&&\times \left(\frac{4m^2}{t}\right)^{n_2}\frac{%
(-i|n_1)(-k|n_2)\left.\left(\frac{1-i+j-k+l}{2}\right|n_1+n_2\right)}
{(1-i+j|n_1)(1-k+l|n_2)(1+j+l+\frac{D}{2}|n_1+n_2)}\nonumber \\
&&\times\left(\left.1+\half (-i+j-k+l)\right|n_1+n_2\right),  
\end{eqnarray}  
and

\begin{eqnarray}
{  I}_5 &=&\frac{2\pi ^{D/2}(-t)^l(-s)^i(-m^2)^{\frac{D}{2}+j+k}(-j|i)
(-k|l)}{(1-i+j+k-l|\frac{D}{2}+i+l)}\sum_{n_1,n_2=0}^\infty \frac 1 
{n_1!n_2!}\left( \frac{4m^2}s\right) ^{n_1}  \nonumber \\
&&\times\left( \frac{4m^2}t\right) ^{n_2}\frac{%
(-i|n_1)(-l|n_2)\left.\left(\frac{1-i+j+k-l}{2}\right|n_1+n_2\right)}{
(1-i+j|n_1)(1+k-l|n_2)(1+j+k+\half D|n_1+n_2)}\nonumber \\
&&\times\left.\left(1+\half (-i+j+k-l)\right|n_1+n_2\right)  
\end{eqnarray}

From these we construct the second hypergeometric series representation
for the Feynman integral as just the linear combination
$I_2+I_3+I_4+I_5$, i.e., 

\begin{eqnarray}
J_2(i,j,k,l;m) &=&\frac{2\pi ^2}{st}\sum_{\{a,b,c,d\}}\frac{\Gamma
(c-a)\Gamma (d-b)}{\Gamma (c)\Gamma (d)} \label{J2} \\
&&\times F_2\left( 1+a+b-\gamma ,a,b;1+a-c,1+b-d\left|
\frac{4m^2}s,\frac{4m^2}{t}\right) \right.\nonumber\end{eqnarray}
where the set $\{a,b,c,d\}$ takes the values $\{-i,-k,-j,-l\}$, 
$\{-i,-l,-j,-k\}$, $\{-j,-k,-i,-l\}$, $\{-j,-l,-i,-k\}$ and $F_2$ 
is another hypergeometric function\cite{bateman,appel}. 

Note that $J_2$ is the \ca{} of $J_1$, see for example Erd\'elyi 
{\it et al} \cite{bateman}. While $J_1$ is valid in the
non-relativistic  case and $J_2$ in the relativistic one, they do not
cover all the possible regions of external momenta. 

An important point here that one must be aware of is that even though
singularities might appear in isolated terms of the RHS, for the
special case when $i=j=k=l=-1$, the above equation (\ref{J2}) cannot be
singular since the corresponding \ca{} formula (\ref{J1}) is not.

To overcome this difficulty let us introduce small corrections
in the parameters $\beta $ and $\beta'$ of the hypergeometric
function $F_3(\alpha,\alpha',\beta,\beta'; \gamma|x,y)$\cite{davyd}. This
recourse corresponds to correcting the exponents of
propagators\cite{letb}. In our case we take 
\[
\beta \rightarrow 1+\delta\,\,,
\]
\[
\beta ^{\prime}\rightarrow 1+\delta ^{\prime }.
\]

Next we expand all the (gamma) factors which contain $\delta$ and 
$\delta^{\prime}$ and the $F_2$ functions in Taylor series around 
$\delta=0$ and $\delta'=0$. In the end we take the limit of vanishing 
$\delta$ and $\delta^{\prime}$. This result is valid, like in the first 
case, within the region of convergence of the series which defines the 
$F_2$ function\cite{bateman,appel},
\[
F_2(\alpha ,\beta ,\beta ^{\prime };\gamma ,\gamma ^{\prime
}|z_1,z_2)=\sum_{j,k=0}^\infty \frac{z_1^jz_2^k}{j!k!}\frac{(\alpha
|j+k)(\beta |j)(\beta ^{\prime }|k)}{(\gamma |j)(\gamma ^{\prime }|k)}%
,\quad |z_1|+|z_2|<1 
\]

Following these steps we arrive at Davydychev's second expression for
the Feynman integral\cite{stand}. Note that now the Mandelstam's
variables $s$ and $t$ can never be zero. Moreover, like the case of
$J_1$, he has shown that the resulting expression for $J_2$ can be
converted into the standard result in terms of sum of logarithms and
dilogarithms through the use of a double integral representation for
$F_2$\cite{bateman}. 

We see that with NDIM we {\em immediately} get two \hf{} which are
related by \ca{}. But as we have mentioned before, there are still ten
other solutions for the system, so that we expect that these new
results will generate solutions for the integral that are also related
by \ca{}. The theory of generalized \hf{}\cite{bateman,edinburgo} tells
us that there are three \an{} continuations with no restriction in the
parameters, namely: 
\[
F_3(...|x,y) \longrightarrow F_2(...|1/x,1/y). \]

This is the very relation we saw above, connecting the two \hf{}. The
other 
two,
\[ F_3(...|x,y) \longrightarrow H_2(...|1/x,-y) \]
and
\[ H_2(...|x,y) \longrightarrow F_2(...|1/x,-y) \]
must appear in our new results whether in the form of direct or
indirect  \an{} continuation. So these new results for the Feynman
integral (\ref{K}) will cover all the possible physical regions of
external momenta.

\section{Particular Cases.} 

If we put in our integral any exponent equal to zero we get an integral 
with three massive propagators. NDIM must be consistent for any value
of the exponents, so, taking for example $i=0$ the result (\ref{I1})
gives 

\begin{eqnarray}  \label{3p1}
J_1(0,j,k,l;m) &=& {  C}^{(1)}\;\\
&&\times\; _3F_2 \left(\matrix{\!\!\!\!\!\!-k,& 
\!\!\!\!\!\!\!\!\!\!\!\!\!\!\!\!\!-l,&
\!\!\!\!\!\!\!\!\!\!\!\!-j-k-l-\frac{D}{2},
\cr\quad\;\;\;\;\;\;\frac{-j-k-l} {2},&
\quad\;\;\;\;\;\;\frac{1-j-k-l}{2}}\right.\left|\frac{t}{4m^2} 
\right)  \nonumber \end{eqnarray}
where $_3F_2$ is the generalized hypergeometric function of one 
variable\cite{bateman} and 
\[
{  C}^{(1)} =
\left(\frac{-\pi}{2}\right)^{D/2}\frac{2\sqrt{\pi}(-2m^2)^  
{j+k+l+\frac{D}{2}}\Gamma(-j-k-l-\frac{D}{2})}{\G(\frac{-j-k-l}{2}) 
\Gamma(\frac{1-j-k-l}{2})} 
\]
The same result arises if one puts $j=0$. For $k=0$ we need simply to
replace $ s\leftrightarrow t$ to get

\begin{eqnarray}  \label{3p2}
J_1(i,j,0,l;m) &=& {  C}^{(2)}\;\\ 
&&\times \;_3F_2 \left(\left.\matrix{-i, & -l,& -i-j-l-\frac{D}{2},
\cr\quad  \frac{-i-j-l}{2}, &
\quad\frac{1-i-j-l}{2}}\right|\frac{s}{4m^2}\right) 
\nonumber\end{eqnarray}
where 
\[
C^{(2)}=\left(\frac{-\pi}{2}\right)^{D/2}\frac{2\sqrt{\pi}(-2m^2)^ 
{i+j+l+\frac{D}{2}}\G(-i-j-l-\frac{D}{2})} {\Gamma(\frac{-i-j-l}{2})
\Gamma(\frac{1-i-j-l}{2})} 
\]
Finally, if we take $l=0$ we get the same result as above in virtue of
the symmetry ($k\leftrightarrow l$) in equation (\ref{I1}) .

From these equations above we can also calculate integrals with 
two massive propagators, i.e., vacuum polarization-like ones. In
(\ref{3p1}) taking for example $j=0$,we get

\[
J_1(0,0,k,l;m)=C^{(3)}\; _3F_2 \left(\left.\matrix{-k,& -l,& -k-l- 
\frac{D}{2}, \cr\quad\frac{-k-l}{2},& \quad\frac{1-k-l}{2}}\right| 
\frac{t}{4m^2}\right) \]
where 
\[
{  C}^{(3)}= \left(\frac{-\pi}{2}\right)^{D/2}\frac{2\sqrt{\pi}
\Gamma(-k-l-\frac{D}{2})} {(-2m^2)^{-k-l-\frac{D}{2}} 
\Gamma(\frac{-k-l}{2}) \Gamma(\frac{1-k-l}{2})} 
\]
In an analogous way, if we take $l=0$ in (\ref{3p2}),we have

\[ J_1(i,j,0,0;m) = C^{(4)}\; _3F_2 \left(\left.\matrix{-i,& -j,& -i-j-
\frac{D}{2}, \cr\quad\frac{-i-j}{2},& \quad\frac{1-i-j}{2}}\right| 
\frac{t}{4m^2}\right)\]
where 
\[
C^{(4)} = \left(\frac{-\pi}{2}\right)^{D/2}\frac{2\sqrt{\pi}
\G(-i-j-\frac{D}{2})}{(-2m^2)^{-i-j-\frac{D}{2}}\G(\frac{-i-j}{2}) 
\G(\frac{1-i-j}{2})} 
\]

For the particular cases where the exponents are equal to minus one the
results above are respectively given by,

\begin{equation}
J_1(-1,-1,0,-1;m)=\frac{-\pi ^2}{2m^2}{\;_3F_2}\left( \left. 
\matrix{1,& 1,& 1, \cr\quad 2,& \quad\frac{3}{2}}\right| 
\frac s{4m^2}\right) \end{equation}
which can be written in terms of elementary functions\cite{boos}. Here
we have already taken the limit $D=4$, since the result is finite in
this limit, and 

\begin{equation}
J_1(0,0,-1,-1;m)=\pi ^2(m^2)^{2-\half D}\Gamma
(2-\frac{D}{2})\;_2F_1\left( \left.  
\matrix{1,& 2-\frac{D}{2} \cr& \quad\frac{3}{2}}\right| 
\frac s{4m^2}\right) 
\end{equation}
where now, of course, there is a well-known single pole in the limit
$D=4$. 

If one is interested in other regions of external momenta one has to 
take the \ca{}\cite{luke} of these results. We do not need to start
from the very beginning, eq.(\ref{I1}), to construct them in other
regions,  because we know from the theory of complex variables that any
functional property of one power series is shared by all the
others\cite{kaplan}. 

\section{Conclusion.} 

Using NDIM we have evaluated a massive box diagram integral, namely, a
Feynman integral bearing four massive propagators. This integral is the
one appearing in the \yy{} in QED and the two well-known results,
expressed in terms of hypergeometric functions, have been easily found.
So, the computation of such an integral, done as a ``lab test'' for
NDIM, has revealed to us a powerful technique, which transfers the
intricacies of performing Feynman integrals in positive dimensions to
that of solving a system of linear algebraic equations in negative
dimensions, a far simpler task to perform than, e.g., solving
parametric integrals. More than that, surprisingly, the technique not
only reproduces the standard results, but gives simultaneously,
solutions covering other regions of the external momenta. We are
studying carefully such solutions and these are going to be the subject
of our shortly forthcoming paper. Also more complicated box diagram
integrals, such as arising from the one-loop correction to the
Bhabha-scattering in QED and an off-shell two-loop triangle diagram
integral have already been checked by us.

\begin{ack}
AGMS would like to thank Prof. Andrei Davydychev for helpful hints and
very clear discussions of \cite{stand}. The work of AGMS is supported
by CNPq (Conselho Nacional de Desenvolvimento Cient\'{\i}fico e
Tecnol\'ogico, Brazil). 
\end{ack}

\end{document}